\documentclass[showpacs, pra,twocolumn,preprintnumbers ,amsmath, amssymb, superscriptaddress, aps]{revtex4-2}
\usepackage{color}
\usepackage{amsmath,amssymb}
\usepackage{pifont}
\usepackage{amssymb}  
\usepackage{bbold}
\usepackage{float}
\usepackage{subfloat}
\usepackage{adjustbox}

\usepackage[caption=false]{subfig}
\usepackage{tikz}
\usepackage{makecell}
\usepackage{subfig}
\usepackage{pifont}   
\usepackage{graphicx} 
\graphicspath{{Figures/}}
\usepackage{dcolumn}  
\usepackage{bm}       
\usepackage{multirow} 
\usepackage{placeins}
\usepackage[colorlinks]{hyperref}
\usepackage{mathtools}
\usepackage{appendix}

\newcommand{\lb} {\label}
\newcommand{\bb} {\color{blue}}

\def \be{\begin{align}}
	\def \ee{\end{align}}
\def \bea{\begin{eqnarray}}
	\def \eea{\end{eqnarray}}

\begin{document}
	
	\title{Transmissions in gapped graphene exposed to tilting and oscillating barriers}
	\author{Miloud Mekkaoui}
	\affiliation{ Laboratory of Theoretical Physics, Faculty of Sciences, Choua\"ib Doukkali University, PO Box 20, 24000 El Jadida, Morocco}	
	\author{Ahmed Jellal}
	\affiliation{ Laboratory of Theoretical Physics, Faculty of Sciences, Choua\"ib Doukkali University, PO Box 20, 24000 El Jadida, Morocco}
	\affiliation{
		Canadian Quantum  Research Center,
		204-3002 32 Ave Vernon, Vernon BC V1T 2L7,  Canada}
	\author{Abderrahim El Mouhafid}
	\affiliation{ Laboratory of Theoretical Physics, Faculty of Sciences, Choua\"ib Doukkali University, PO Box 20, 24000 El Jadida, Morocco}

\begin{abstract}
 We investigate the transmissions of fermions through gapped graphene
	structures by employing a combination of double barrier tilting and a time-oscillating potential. The latter introduces additional sidebands into the transmission probability, which manifest at energy levels determined by the frequency and incident energy. These sidebands arise from the absorption or emission of photons generated by the oscillating potential. We demonstrate that the tilting and positioning of the scattering events within the barriers play a crucial role in determining the peak of tunneling resistance. In particular, the presence of a mid-barrier-embedded scatter leads to a transition from a peak to a cusp when the incident energy reaches the Dirac point within a barrier. Additionally, we illustrate that introducing a time-varying potential results in transmissions dispersing across both the central band and the sidebands.


\end{abstract}


\maketitle


\section{ Introduction}


Graphene \cite{Novoselov1, Geim, Castro} can be modified in different ways such as using a gate voltage, cutting it into nanoribbons, doping, or creating a magnetic barrier. The effects of various types of barriers on transmission and conductance in graphene 
 have been studied, including electrostatic \cite{Ramezani5, Jellal10}, magnetic \cite{Choubabi8,Bahlouli9},  linear \cite{Bahlouli11,Mekkaoui7, Mekkaoui8}, and triangular \cite{Mouhafid12,Mekkaoui6} barriers. 
As for the barrier oscillating in time with frequency $\omega$, it is shown that the tunneling effect exhibited transmissions of additional sidebands at energies $\epsilon + l\hbar \omega$ ($l=0,\pm1,\cdots$).
During this process, energy quanta are exchanged between electrons and photons in the oscillating field. The standard model that accounts for this phenomenon includes a scalar potential that is time-modulated and limited to a finite spatial region. Such results have been proven experimentally by considering photon-assisted tunneling in superconducting films subjected to microwave fields \cite{Dayem} and theoretically by supposing that a microwave field generates a time-harmonic potential difference between the two films  \cite{Tien}. Subsequently, numerous research groups conducted further theoretical studies on this topic and in particular the important case of traversal time of particles that interact with a barrier of time-oscillating \cite{Buttiker}. It has been demonstrated that by employing the transfer matrix, photons can be exchanged between the oscillating potential and electrons, resulting in the transfer of electrons to the sidebands with a certain probability \cite{Wagner}.


In light of the increasing interest in exploring the optical characteristics of electron transport in graphene under intense laser fields, there has been a notable upswing in theoretical investigations on how time-varying periodic electromagnetic fields influence the electronic properties  \cite{Jiang1}.
According to \cite{Calvo}, the electron density of states, as well as the associated electron transport characteristics, can be affected by laser fields.  Laser irradiation-induced electron transport in graphene was found to result in subharmonic resonant enhancement \cite{San-Jose,Kibis2010,Kibis2016, Iurov2020}.
In a recent study \cite{Savelev-1}, an analogy was established between the energy spectra of Dirac fermions in laser fields and those observed in graphene superlattices that are generated from the application of a static one-dimensional periodic potential. It is found that a scalar potential barrier varying in space and time in graphene can enhance electron backscatterings and currents \cite{Savelev-2, Liu}. Applying an oscillating field can lead to the emergence of an effective mass (dynamic gap) \cite{Fistul}. Also adiabatically pumped fermions in graphene subjected to two oscillating barriers studied in \cite{Evgeny}. Furthermore, the transmission probabilities of Dirac fermions in graphene under the influence of a time-oscillating linear barrier potential were studied by us in \cite{Mekkaoui3}.

Various experimental methods exist for inducing a gap in the band structure of graphene, referred to as the Dirac gap \cite{3333}. Experimental evidence illustrates that the maximum energy gap, reaching 260 meV, can result from sublattice symmetry breaking \cite{2525}. It is worth noting that the specific value of the energy gap is subject to variation based on the experimental technique employed. One approach to generating a gap involves manipulating the interface structure between graphene and Ru (ruthenium) \cite{2626}. In the case of graphene epitaxially grown on a SiC substrate, an observable energy gap has been measured \cite{2525}. A gap can be also induced by placing  graphene on  h-BN (hexagonal boron nitride) substrate \cite{8888,9999}. 
The substrate h-BN continues to be an optimal choice for graphene due to its ability to enhance the performance of graphene devices, particularly in challenging conditions such as higher temperatures and higher electric fields
\cite{Wang2017}.

 Creating time-oscillating tilting barriers and a gap in a real experimental setting typically involves the use of external fields or potentials to modulate the potential landscape experienced by particles or waves. One common method is to use an optical lattice, which is created by interfering laser beams to form a periodic potential \cite{1100,2200}. The lattice potential can be tilted by applying a force or a phase shift to the laser beams, leading to a time-oscillating tilting barrier. This can be achieved by modulating the amplitude or phase of the laser beams using acousto-optic modulators or electro-optic modulators.
Another method involves using time-dependent electric fields to create a time-oscillating tilting barrier. This can be achieved by applying a time-varying voltage to electrodes in a setup such as a semiconductor heterostructure or a microfabricated device \cite{3300,4400}. The electric field can be modulated sinusoidally or with other time-dependent waveforms to create the desired time oscillations.

In this study, we extend the outcomes achieved in our previous work \cite{Mekkaoui3} to encompass the case of double barrriers tilting. Specifically, we examine a monolayer graphene sheet placed on the $xy$-plane and exposed to double tilted potential barriers and a time-varying potential in the $x$-direction while the carriers are unrestricted in the $y$-direction. The height of the barrier undergoes sinusoidal oscillations around an average value with an oscillation amplitude and a frequency. We compute the transmission probabilities sidebands, which relies on incident energy, incident angle, and potential parameters. The constraint to near sidebands results from computational challenges in truncating the resulting coupled channel equations, which confines us to low quantum channels. This will allow us to analyze the behavior  of the system under consideration and underline its basic features.

 We highlight that the study of fermion transmissions in gapped graphene structures is motivated by a desire to explore new avenues in the field. By considering a time-oscillating potential, leading to additional sidebands, we provide a novel perspective on how transmission properties can be influenced in such structures. By demonstrating the critical role played by the tilting and positioning of scattering events within the barriers, we aim to contribute to a deeper understanding of the factors influencing the tunneling effect. As a novelty of results,	our investigation reveals a distinctive transmission phenomenon through gapped graphene structures. By incorporating double barrier tilting and a time-oscillating potential, we introduce additional sidebands into the transmission probability. These sidebands emerge at energy levels dictated by the frequency and incident energy, representing the absorption or emission of photons generated by the oscillating potential. The specific interplay between these double barriers  results in unique transmission properties and behaviors that have not been reported in prior works.

The paper is structured in the following manner. Sec. \ref{TTMM} introduces the mathematical background needed to achieve our goal, which includes the system Hamiltonian together with the applied potentials made of double tilting and time-oscillating barriers. These will be used to determine the energy spectrum for each of the five regions composing the system.  Numerical analysis of our results   and comparisons with previously published works are presented in Sec. \ref{DDDD}. Finally,  we conclude our work. In Appendix \ref{TTTT}, we apply  the boundary conditions and use the current density to precisely get the transmissions for all sidebands.


\section{Mathematical tools}\label{TTMM}

We consider Dirac fermions in graphene through  symmetric double barriers tilting subjected to a gap and time-oscillating potential. We use the geometry, depicted in Fig. \ref{db.1},  made of five regions  $j$ = $1, \cdots, 5$: $(1,5)$ contain intrinsic graphene, $(2,4)$ exposed to a linear potential $V(x)$, and 3 subjected to an energy gap together with an oscillating potential around $V_{2}$ with amplitude $U$ and frequency $\omega$. At the interface $x=-L_2$, fermions are incident with energy $E$ and an angle $\phi_{0}$, which quiets the barrier with  $E+ m\hbar \omega$ $(m=0, \pm 1, \pm 2, \cdots)$, forming the angles $\pi-\phi_{m}$ after reflection and $\phi_{m}$ after transmission. 

\begin{figure}[ht] \centering
	\includegraphics[scale=0.6]{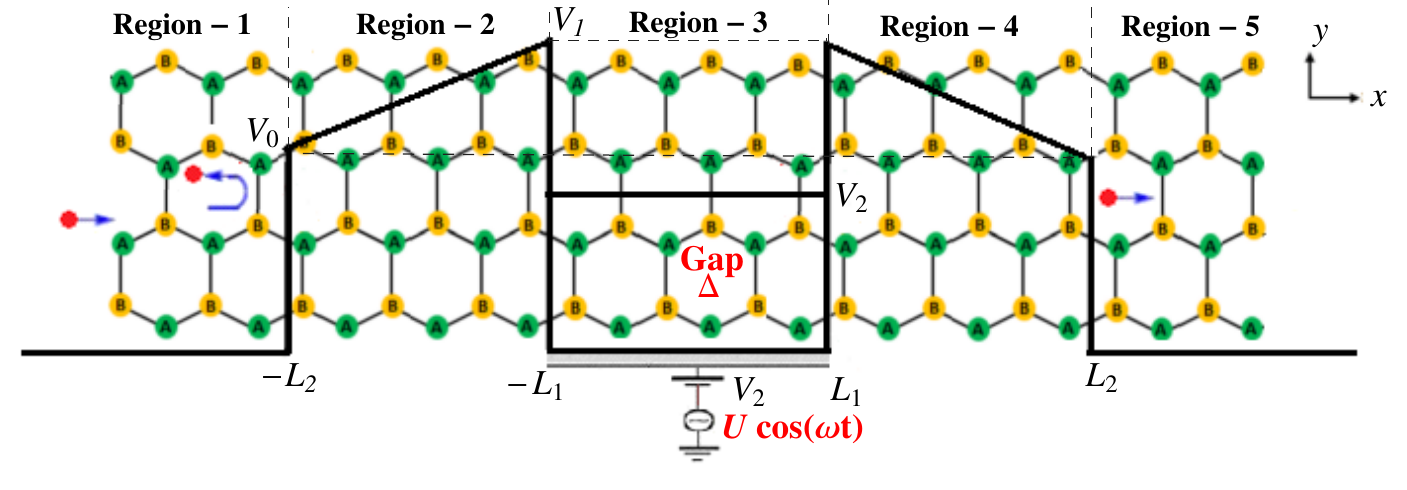}
	\caption{(color online) Representation of Dirac fermions in  gapped graphene through double barriers tilting of widths $L_{2}-L_{1}$ and heights 	$(V_{0}, V_{1})$, in the presence of	
		a time-oscillating potential $U\cos(\omega t)$ of width $2L_{1}$, amplitude $U$ and frequency $\omega$. }\lb{db.1}
\end{figure}

To provide a description of the present system, we use the Hamiltonian for one Dirac point, e.g., $ K$,
\begin{equation}\lb{eq1}
H_j=v_{F} {\boldsymbol{\sigma}}\cdot\textbf{p}+ V_{j}(x){\mathbb
I}_{2}+\left[\Delta\sigma_{z}+U\cos(\omega
t){\mathbb
	I}_{2}\right]\Theta\left(L_{1}^{2}-x^{2}\right)
\end{equation}
where   ${v_{F}\approx 10^6}$ m/s  is the Fermi velocity,  $\textbf{p}=-i\hbar(\partial_{x},\partial_{y})$, ${{\boldsymbol{\sigma}}=(\sigma_{x},\sigma_{y})}$ are the Pauli
matrices, ${\mathbb I}_{2}$ the $2 \times 2$ unit matrix, and 
$\Theta$ is the Heaviside step function.
 $\Delta = {m_0} v_{F}^2$ represents the energy gap that results from the sublattice symmetry breaking, or that  arising from the spin-orbit interaction $\Delta = \Delta_{so}$. The electrostatic potential {$V_{j}(x)$} in each  region $j$ is expressed as follows:
\begin{equation}\lb{eq2}
{V_{j}(x)}=
\left\{%
\begin{array}{ll}
    \gamma F x+V_L, & \hbox{$L_{1}\leq |x|\leq L_{2}$} \\
    V_{2}, & \hbox{$ |x|\leq L_{1}$} \\
    0, & \hbox{otherwise} \\
\end{array}%
\right.
\end{equation}
with $F=\frac{V_1-V_0}{L_2-L_1}$, $V_L=\frac{V_1L_2-V_0L_1}{L_2-L_1}$, 
$\gamma=1$ stands for $x\in [-L_2, -L_1]$, and 
$\gamma=-1$ for $x\in [L_1, L_2]$.
 The eigenvalue  equation for the spinor $\psi_{j}(x,y,t)$ at $E+l\hbar\omega$, in the unit system $\hbar=1$, reads as
\begin{widetext}
	\begin{equation}\lb{eq4}
\left[{\boldsymbol{\sigma}}\cdot\textbf{p}+ v_{j}(x){\mathbb
I}_{2}+\left(\Delta_{g}\sigma_{z}+u\cos(\omega
t){\mathbb
	I}_{2}\right)\Theta\left(L_{1}^{2}-x^{2}\right)\right]\psi_{
j}(x,y,t)=(\epsilon+l\varpi)\psi_{j}(x,y,t)
\end{equation}
\end{widetext}
and we have define the quantities $\epsilon=\frac{E}{
\upsilon_{F}}$, $\varpi=\frac{\omega}{ \upsilon_{F}}$,
$v_j=\frac{V_j}{v_F}$, $u=\frac{U}{v_F}$,
$\Delta_{g}=\frac{\Delta}{v_F}$ and $\alpha=\frac{u}{\varpi}$. 
The system is expected to exhibit finite width W and infinite mass boundary conditions at the boundaries $y=0$ and $y=W$ \cite{Tworzydlo, Berry}, resulting in a quantized  wave vector along the $k_y=k_n$
\begin{equation}
k_n=\frac{\pi}{W}\left(n+\frac{1}{2}\right),\quad n\in \mathbb{N}
\end{equation}

 By considering energy conservation, we can represent the wave packet of electrons in the $j$-th region using a linear combination of wave functions that have energies $\epsilon+m\omega$. Then, after solving the eigenvalue equation, we get  the eigenspinor for  incident and
reflection waves in region $j$=1 ($x \leq -L_2 $)
\begin{widetext}
	\begin{equation}
\psi_{
1}(x,y,t)=e^{ik_{y}y}\sum^{+\infty}_{m,l=-\infty}\left[\delta_{l,0}\left(
\begin{array}{c}
1 \\
 z_{l}\end{array}\right)e^{ik_{l}x}+r_{l}\left(
\begin{array}{c}
1 \\
 -\frac{1}{z_{l}}\end{array}\right)e^{-ik_{l} x
 }\right]\delta_{m,l}e^{-iv_{F}(\epsilon+m\varpi)t}
\end{equation}
\end{widetext}
where $z_{l}=s_{l}\frac{k_{l} +ik_{y}}{\sqrt{k^{2}_{l}
+k_{y}^{2}}}$, $s_{l}=\mbox{sgn}(\epsilon+l \varpi)$,  and
\begin{align}
	k_{l}=s_{l}\left(\left(\epsilon+l\varpi\right)^{2}-k^{2}_{y}\right)^{\frac{1}{2}}.
\end{align}
In the transmitted
region $j$=5 ($ x \geq L_2 $), we obtain the solution
\begin{widetext}
	\begin{equation}
\psi_{
5}(x,y,t)=e^{ik_{y}y}\sum^{+\infty}_{m,l=-\infty}\left[t_{l}\left(
\begin{array}{c}
1 \\
 z_{l}\end{array}\right)e^{ik_{l} x
 }+b_{l}\left(
\begin{array}{c}
1 \\
 -\frac{1}{z_{l}}\end{array}\right)e^{-ik_{l} x}\right]\delta_{m,l}e^{-iv_{F}(\epsilon+m\varpi)t}
\end{equation}
\end{widetext}
where $\{b_{l}\}$ is the null vector. In region $j$=3 ($-L_1 \leq
x \leq L_1 $), the solution is
\begin{widetext}
	\begin{eqnarray}
\psi_{3}(x,y,t)
&=&e^{ik_{y}y}\sum^{m,l=+\infty}_{m,l=-\infty}\left[a_{l}^{j}\left(
\begin{array}{c}
\lambda_{l}^{+} \\
 \lambda_{l}^{-} z_{l}^{'}\end{array}\right)e^{ik^{'}_{l}x}+b_{l}^{j}\left(
\begin{array}{c}
\lambda_{l}^{+} \\
 -\frac{\lambda_{l}^{-}}{z_{l}^{'}}\end{array}\right)e^{-ik^{'}_{l} x}\right]
J_{m-l}\left(\alpha\right)e^{-iv_{F}(\epsilon+m\varpi)t}
\end{eqnarray}
\end{widetext}
where
$\lambda_{l}^{\pm}=\left(1\pm\frac{\Delta_{g}}{\epsilon-v+l\varpi}\right)^{\frac{1}{2}}$,
$z_{l}^{'}=s_{l}^{'}\frac{k_{l}^{'}+ik_{y}}{\sqrt{k_{l}^{'}+k_{y}^{2}}}$,
$s_{l}^{'}=\mbox{sgn}(\epsilon+l \varpi-v)$, $J_{m}
\left(\frac{u}{\varpi}\right)$ is the  Bessel function, and the wave vector is given by
\begin{align}
	k_{l}^{'}=s_{l}^{'}\left(\left(\epsilon-v_{2}+l\varpi\right)^{2}-\Delta_{g}^{2}-{k_{y}}^{2}\right)^{\frac{1}{2}}.
\end{align}
In regions $j$=1, 2, 4, 5 the modulation amplitude is null,
$\alpha=0$, and therefore we have  we have the function $J_{m-l}
(0)=\delta_{m,l}$.

 In regions {2} and {4} ($L_{1}<|x|<L_{2}$), the parabolic cylinder function can be used to represent the general solution \cite{Abramowitz,
Gonzalez, Bahlouli11} 
\begin{equation}\lb{hiii1}
 \chi_{\gamma,l}^{+}=c_{n1}
 D_{\nu_n-1}\left(Q_{\gamma,l}\right)+c_{n2}
 D_{-\nu_n}\left(-Q^{*}_{\gamma,l}\right)
\end{equation}
where $F=v_{F}\varrho$, $\nu_n=\frac{ik_{y}^{2}}{2\varrho}$,
$\epsilon_{l}=\epsilon+l\varpi-v_{1}$ and $ Q_{\gamma,
l}(x)=\sqrt{\frac{2}{\varrho}}e^{i\pi/4}\left(\gamma \varrho
x+\epsilon_{l}\right) $, $c_{n1}$ and $c_{n2}$ are constants. The second spinor component can be derived as
\begin{widetext}
	\begin{eqnarray}\lb{hiii2}
\chi_{\gamma,l}^{-}=-\frac{c_{n2}}{k_{y}}\left[
2(\epsilon_{l}+\gamma \varrho x)
 D_{-\nu_n}\left(-Q^{*}_{\gamma,l}\right)
+
 \sqrt{2\varrho}e^{i\pi/4}D_{-\nu_n+1}\left(-Q^{*}_{\gamma,l}\right)\right]
 -\frac{c_{n1}}{k_{y}}\sqrt{2\varrho}e^{-i\pi/4}
 D_{\nu_n-1}\left(Q_{\gamma,l
 }\right).
\end{eqnarray}
\end{widetext}
Now, by defining the components $\varphi_{\gamma,l}^{+}(x)=\chi_{\gamma,l}^{+}+i\chi_{\gamma,l}^{-}$
and
$\varphi_{\gamma,l}^{-}(x)=\chi_{\gamma,l}^{+}-i\chi_{\gamma,l}^{-}$, yield  the eigenspinors solution of 
\eqref{eq4} in the form provided below
\begin{widetext}
	\begin{eqnarray}
\psi_{j}(x,y,t)
&=&e^{ik_{y}y}\sum^{m,l=+\infty}_{m,l=-\infty}\left[a_{l}^{j}\left(%
\begin{array}{c}
 \eta^{+}_{\gamma,l}(x) \\
  \eta^{-}_{\gamma,l}(x) \\
\end{array}%
\right)e^{ik_{y}y}+b_{l}^{j}\left(%
\begin{array}{c}
 \xi^{+}_{\gamma,l}(x) \\
 \xi^{-}_{\gamma,l}(x)\\
\end{array}%
\right)\right]
J_{m-l}\left(\alpha\right)e^{-iv_{F}(\epsilon+m\varpi)t}
\end{eqnarray}
\end{widetext}
and we have set the functions
\begin{widetext}
	\begin{align}
&\eta^{\pm}_{\gamma,l}(x)=
 D_{\nu_{n}-1}\left(Q_{\gamma,l}\right)\mp
 \frac{1}{k_{y}}\sqrt{2\varrho}e^{i\pi/4}D_{\nu_{n}}\left(Q_{\gamma,l}\right)
\\
&
\xi^{\pm}_{\gamma,l}(x)=
 \pm\frac{1}{k_{y}}\sqrt{2\varrho}e^{-i\pi/4}D_{-\nu_{n}+1}\left(-Q_{\gamma,l}^{*}\right)
  \pm
 \frac{1}{k_{y}}\left(-2i\epsilon_{0}\pm
 k_{y}-\gamma2i \varrho x\right)D_{-\nu_{n}}\left(-Q_{\gamma,l}^{*}\right)
\end{align}
\end{widetext}
where $\gamma=\pm 1$. We will see in the forthcoming analysis how the above mathematical tools can be employed to achieve our goals.

\section{Discussions}\label{DDDD}

\begin{figure}[hbt!]
	\centering
	\subfloat[]{
		\centering
		\includegraphics[scale=0.75]{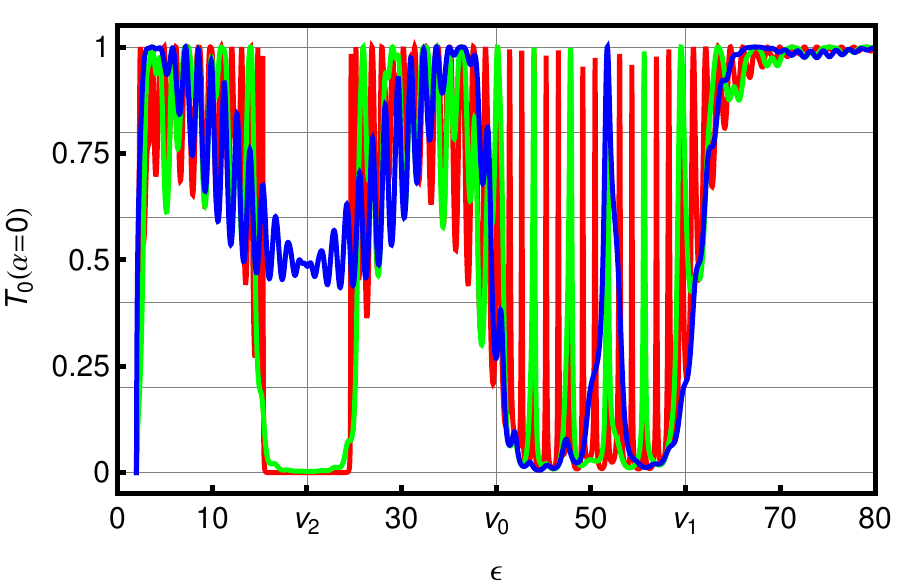}
		\label{fig1a}
	}\\ \subfloat[]{
		\centering
		\includegraphics[scale=0.75]{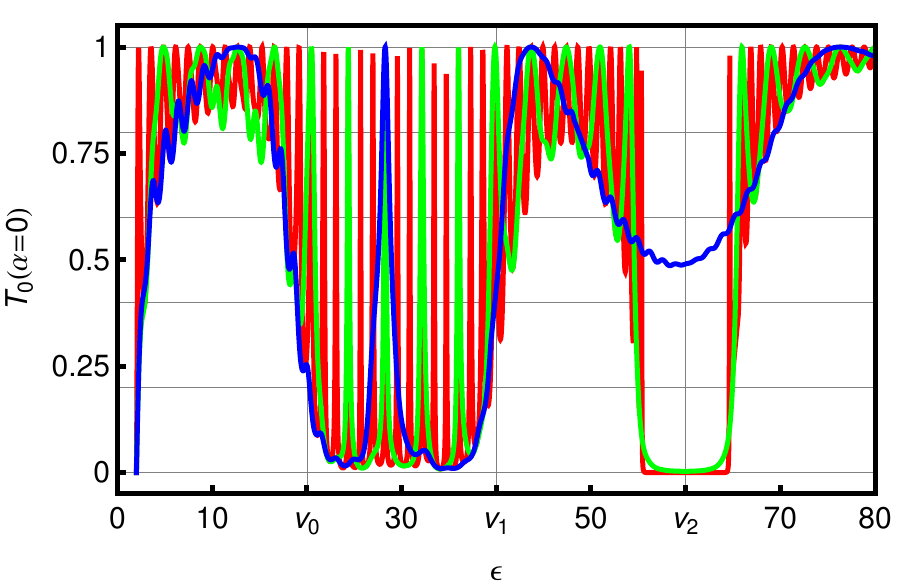}
		\label{fig1b}}
	\\\subfloat[]{
		\centering
		\includegraphics[scale=0.75]{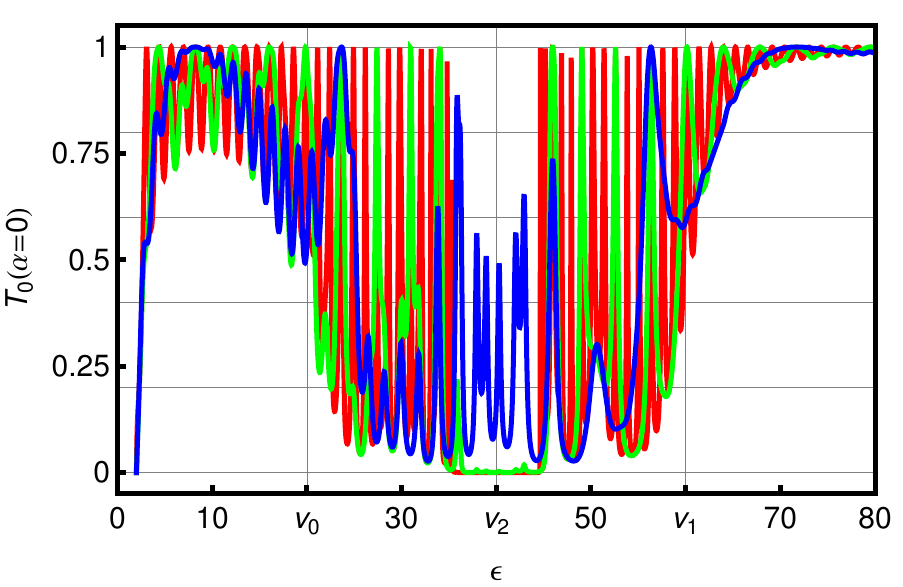}
		\label{fig1c}}
	\caption{{(color online) Transmission    $T_{0}(\alpha=0)$ versus incident energy
			$\epsilon$ for $\Delta_g=4$, $k_{y}=2$, $L_{2}=2.5$, and $L_1=0.1$ (blue line), $L_1=0.4$ (green line),
			$L_1=1.2$ (red line). (a): 
			$v_{0}=40$, $v_{1}=60$, $v_{2}=20$, (b): $v_{0}=20$, $v_{1}=40$,
			$v_{2}=60$, and (c):  $v_{0}=20$, $v_{1}=60$, $v_{2}=40$.}}
	\label{fig1}
\end{figure}

\begin{figure}[ht]
	\centering
	\subfloat[]{
		\centering
		\includegraphics[scale=0.85]{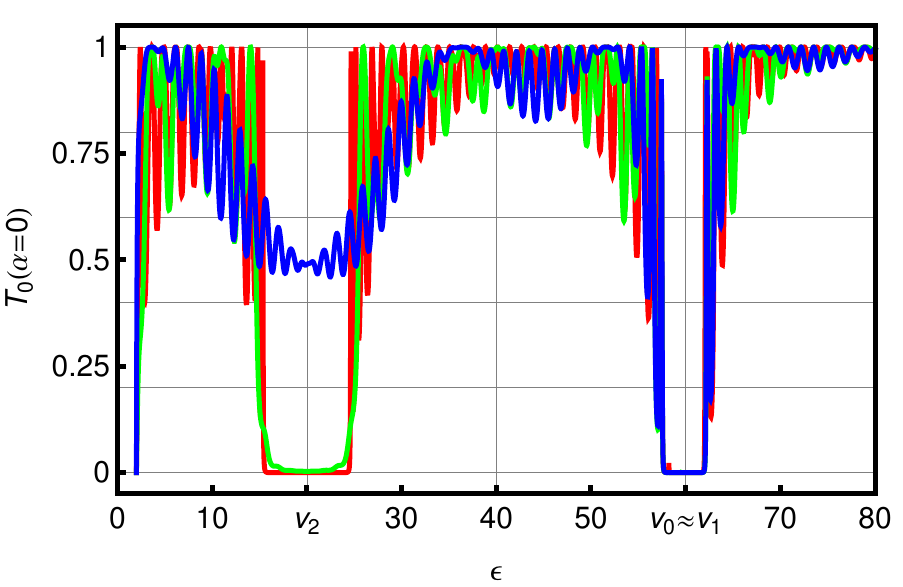}
		\label{fig2a}
	}\\
	\subfloat[]{
		\centering
		\includegraphics[scale=0.85]{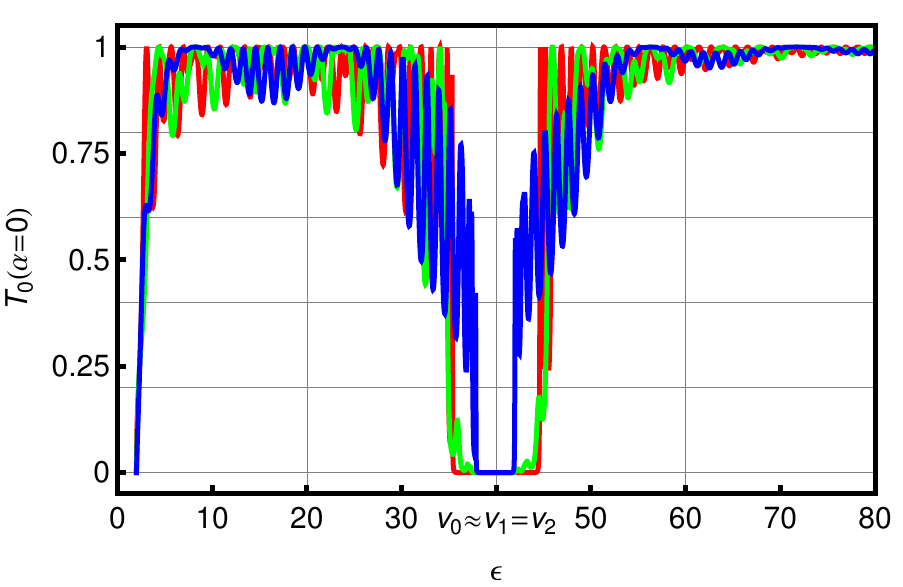}
		\label{fig2b}}
	\caption{{(color online) 
	The same as in Fig. \ref{fig1}	but with	(a): 
			$v_{0}\approx v_{1}=60$, $v_{2}=20$, and  (b): $v_{0}\approx
			v_{1}=v_{2}=40$.}}
	\label{fig2}
\end{figure}

\begin{figure}[ht]
        \centering
        \subfloat[]{
            \centering
            \includegraphics[scale=0.75]{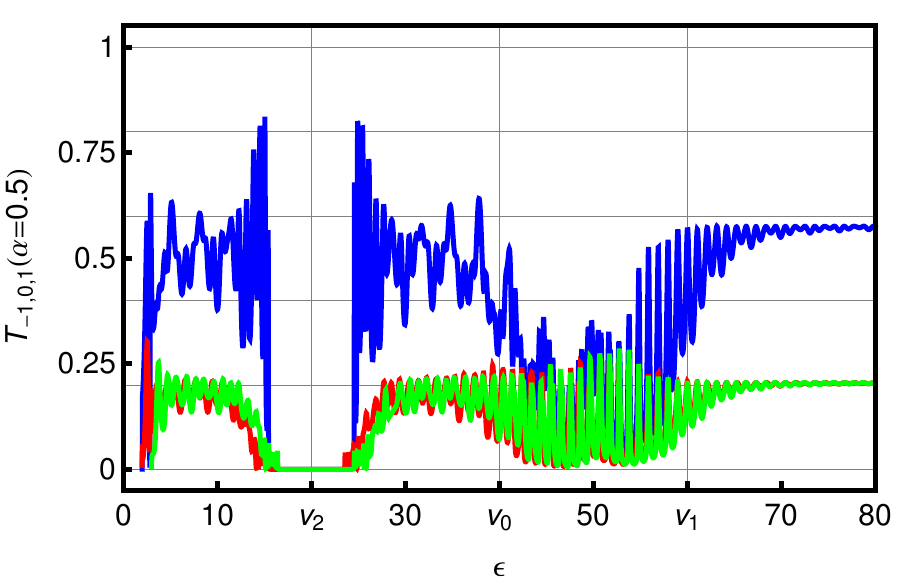}
            \label{ig3a}
        }\\
        \subfloat[]{
            \centering
            \includegraphics[scale=0.75]{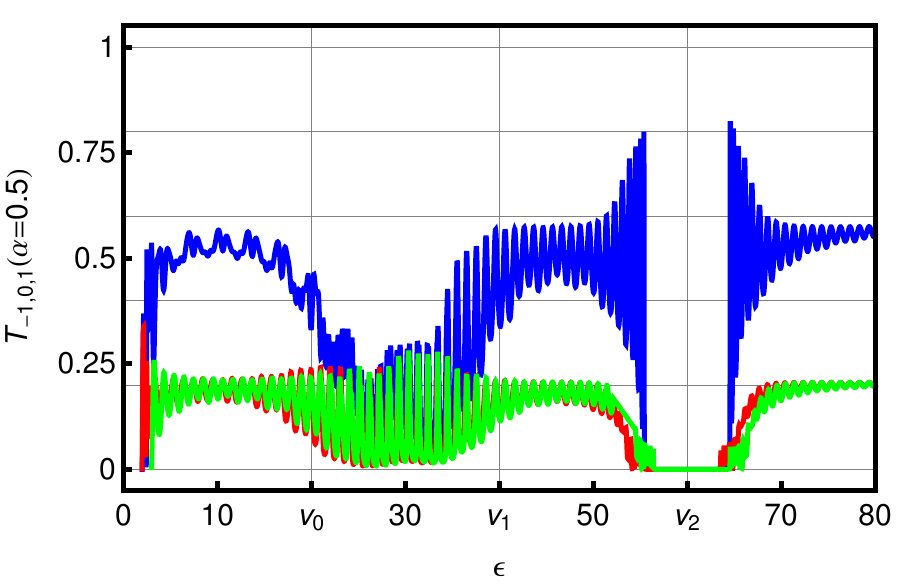}
            \label{fig3b}}\\
            \subfloat[]{
            \centering
            \includegraphics[scale=0.75]{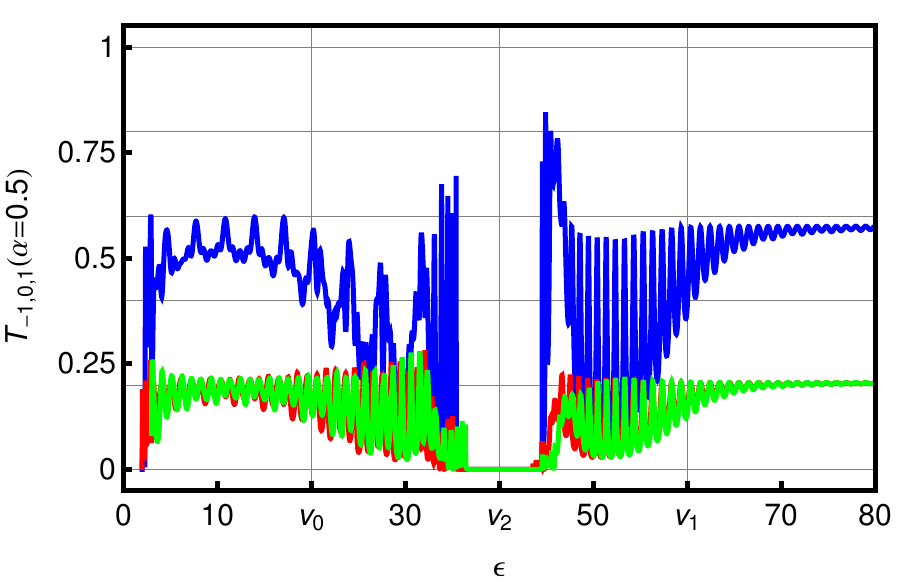}
            \label{fig3c}}
    \caption{{(color online) 
    		Transmissions versus incident energy
 $\epsilon$ for  $\alpha=0.5$, $\Delta_g=4$,  $k_{y}=2$, $L_1=1.5$, and $L_{2}=2.5$. (a): 
$v_{0}=20$, $v_{1}=40$, $v_{2}=60$, (b): $v_{0}=40$, $v_{1}=60$,
$v_{2}=20$, and (c): $v_{0}=20$, $v_{1}=60$, $v_{2}=40$. $T_{-1}$
(green line), $T_{0}$ (blue line) and $T_{1}$ (red line).}}
    \label{fig3}
\end{figure}

\begin{figure}[ht]
	\centering
	\subfloat[]{
		\centering
		\includegraphics[scale=0.75]{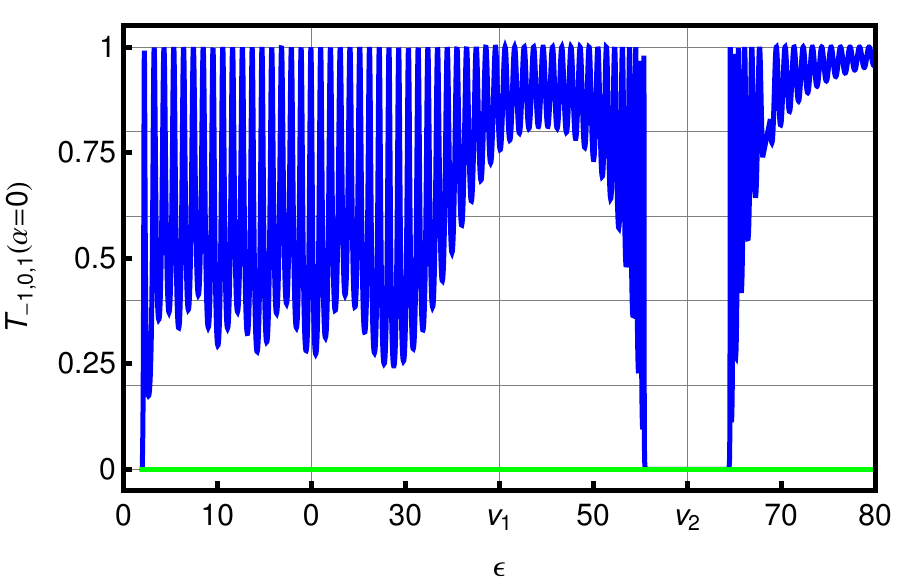}
		\label{fig4a}
	}\\ \subfloat[]{
		\centering
		\includegraphics[scale=0.75]{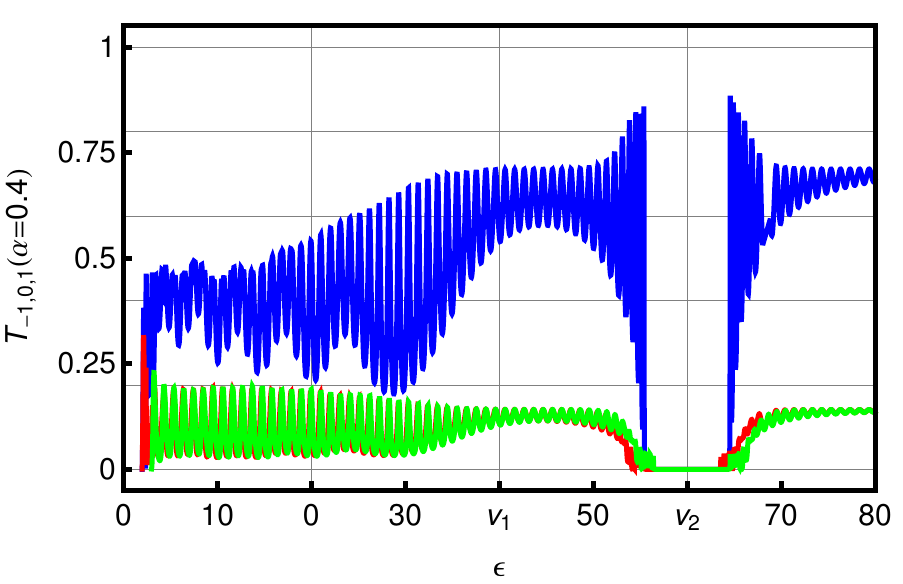}
		\label{fig4b}}\\
	\subfloat[]{
		\centering
		\includegraphics[scale=0.75]{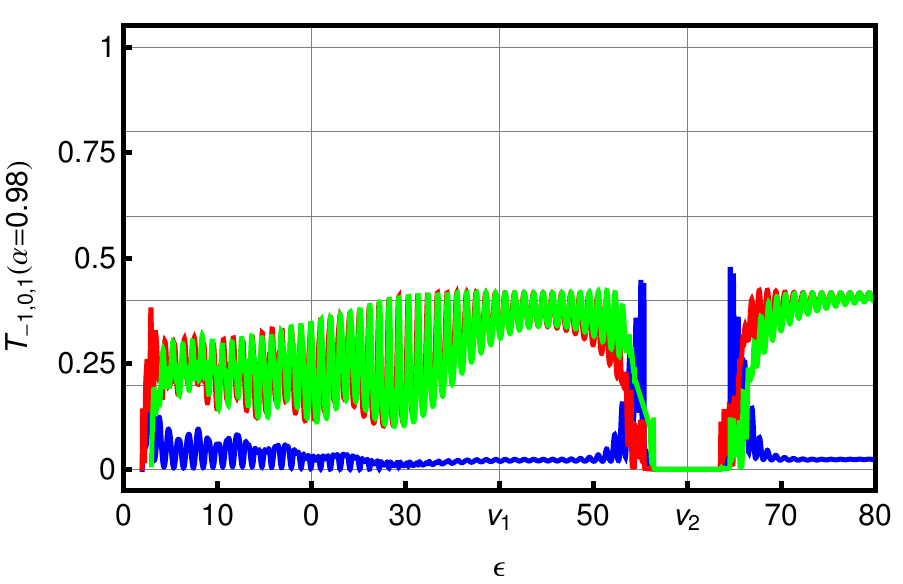}
		\label{fig4c}}
	\caption{{(color online)  
			The same as in Fig. \ref{fig3}	but with $v_{0}=0$, $v_{1}=40$, $v_{2}=60$. (a): $\alpha=0$
			(b): $\alpha=0.4$,  and (c): $\alpha=0.98$.
		}}
	\label{fig4}
\end{figure}

\begin{figure}[ht]
	\centering
	\subfloat[]{
		\centering
		\includegraphics[scale=0.75]{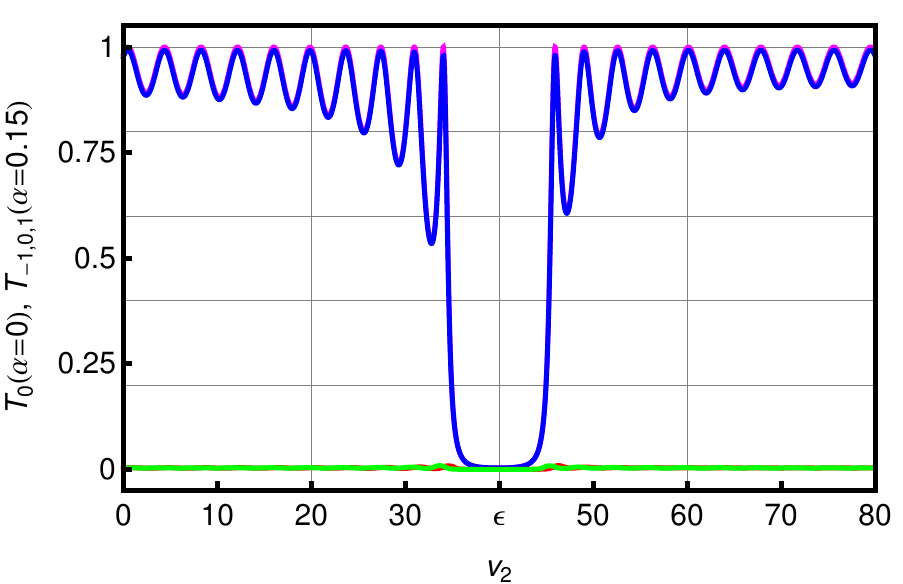}
		\label{fig5a}
	}\\ \subfloat[]{
		\centering
		\includegraphics[scale=0.75]{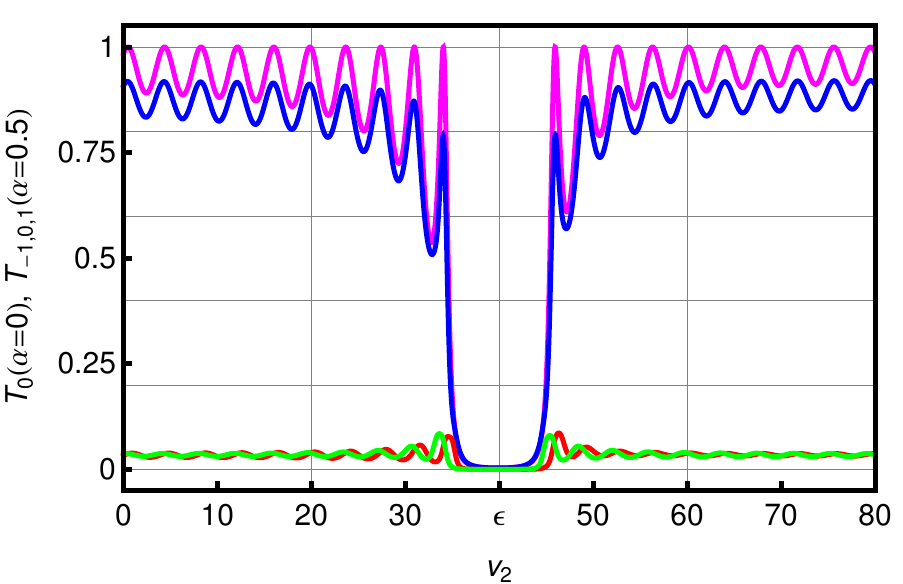}
		\label{fig5b}}\\
	\subfloat[]{
		\centering
		\includegraphics[scale=0.75]{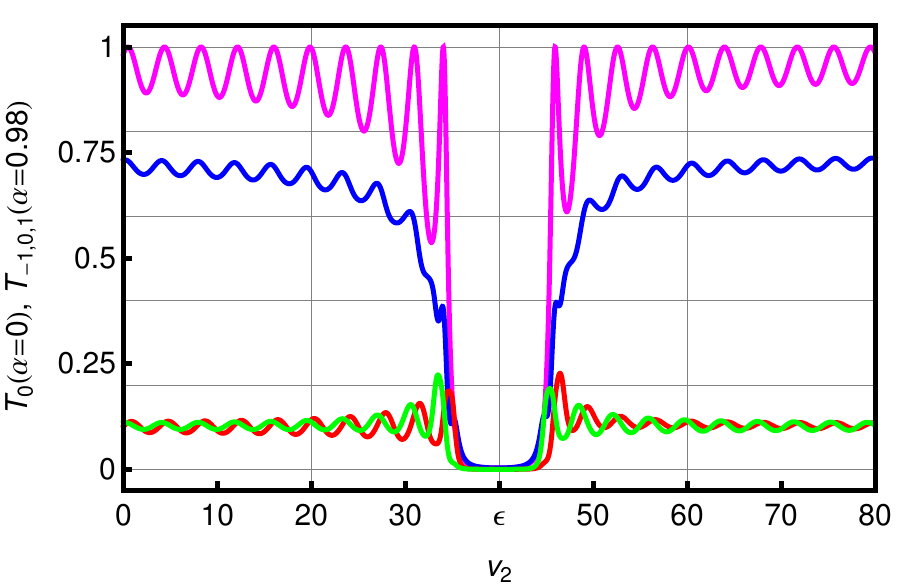}
		\label{fig5c}}
	\caption{{(color online)
			Transmissions versus
			potential $v_2$ for $L_{1}=0.4$, $L_{2}=0.8$, $\Delta_g=4$,
			$k_{y}=2$, $v_{0}=50$, $v_{1}=70$, $\epsilon=60$,
			$T_{0}$ (magenta color), $T_{-1}$ (green line), $T_{0}$
			(blue line), and $T_{1}$ (red line). (a): $\alpha=0.15$, (b):
			$\alpha=0.5$, and (c): $\alpha=0.98$.}}
	\label{fig5}
\end{figure}

\begin{figure}[ht]
        \centering
        \subfloat[]{
            \centering
            \includegraphics[scale=0.85]{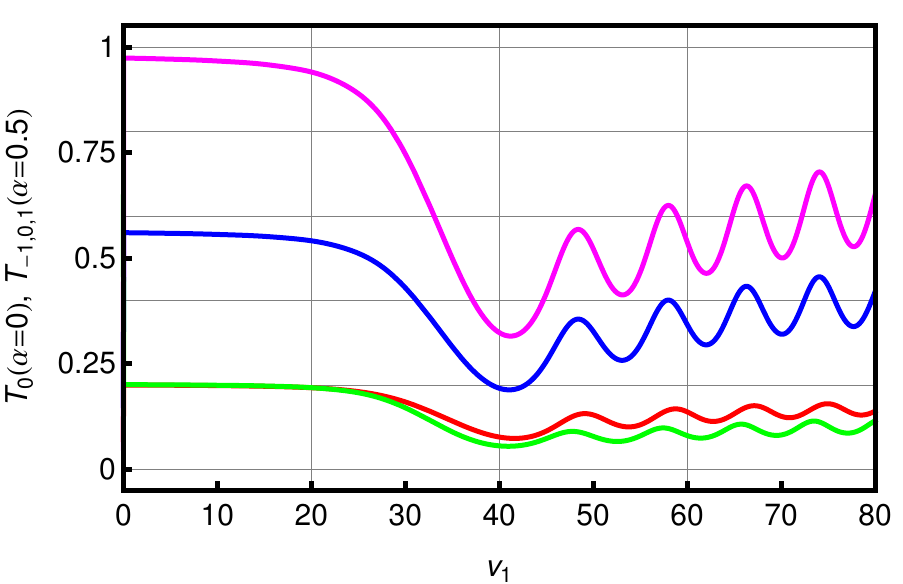}
            \label{fig6a}
        }\\
        \subfloat[]{
            \centering
            \includegraphics[scale=0.85]{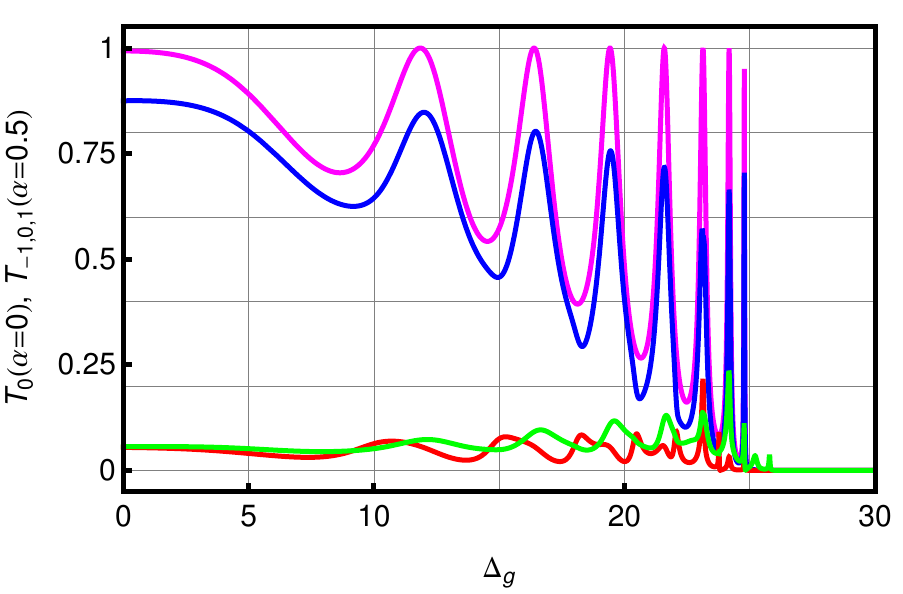}
            \label{fig6b}}
    \caption{{(color online) (a): Transmissions versus
potential $v_1$ for $L_{1}=1.5$, $L_{2}=2.5$, $\Delta_g=4$ and
$k_{y}=2$, $v_{0}=0$, $v_{2}=60$, $\epsilon=30$. (b): Transmissions versus energy gap $\Delta_g$  for $L_{1}=0.5$,
$L_{2}=1.5$, $k_{y}=1$, $v_{0}=0$, $v_1=50$, $v_{2}=40$, and
$\epsilon=15$. $T_{0}(\alpha=0$, magenta line),
$T_{-1}(\alpha=0.5$, green line), $T_{0}(\alpha=0.5$, blue line),
and $T_{1}(\alpha=0.5$, red line).}}
    \label{fig6}
\end{figure}

To illustrate our theoretical results and facilitate numerical calculations, we restrict our analysis to the first three modes: $l=0,\pm 1$, corresponding to the central band and two sidebands, respectively. Consequently, the transmission amplitudes  given in \eqref{ttll} reduce to the following:
\begin{equation}
	t_{-1} = { \mathbb M}_{11}^{-1}[1, 2], \quad t_{0}={ \mathbb M}_{11}^{-1}[2, 2],
	\quad t_{1 }= { \mathbb M}_{11}^{-1}[3, 2].
\end{equation}
At low energies, this approximation can be established because single-photon processes are more likely to occur than two- and higher-photon processes. We will conduct a numerical analysis to better understand the obtained results and highlight the behavior of the system. To accomplish this, we will concentrate on a limited number of transmission channels $(T_0,T_{\pm1})$, see \eqref{ttccs}, and select various configurations of the physical parameters.
The following displays the numerical outcomes for the transmission probabilities. These results are exhibited in Figs. (\ref{fig1}-\ref{fig6}) for multiple parameter values, including $\epsilon$, $v_1$, $v_2$, $\Delta_g$, $L_{1}$, and $L_{2}$.

 Figs. \ref{fig1} and \ref{fig2} illustrate just the transmission probability for the central band $T_{0}$ as a function of energy $\epsilon$ with$\alpha=0$ $L_1=0.1$ (blue line), $L_1=0.4$ (green line), $L_1=1.2$ (red line), $L_{2}=2.5$, $\Delta_g=4$, and $k_{y}=2$. Let us treat double barriers tilting cases when $v_2<v_0<v_1$, $v_0<v_1<v_2$, and $v_0<v_2<v_1$. In both cases, the transmission is plotted in Fig. \ref{fig1}. In Fig. \ref{fig1a}, one can notice that there are six energy zones that characterize the transmission through a graphene double barrier tilting with $v_2<v_0<v_1$. The first is determined by the greater effective mass, namely {\bb $\epsilon<k_y$}, and the second identifies with the lower Klein energy zone characterized by resonances and $k_y<\epsilon<v_2-(k_y+\Delta_g/2)$. Here we have full transmission at some specific energies, despite the fact that the particle energy is less than the height of the barrier. As $L_1$ increases, the oscillations in the Klein zone get reduced. This strong reduction of transmission in the Klein zone seems to suggest the potential suppression of Klein tunneling as we increase $L_1$. The third zone defined by $v_2-(k_y+\Delta_g/2)<\epsilon< v_2+(k_y+\Delta_g/2)$ is a window where the transmission is almost zero. The fourth zone, $v_2+(k_y+\Delta_g/2)<\epsilon<v_0$, is a window where the transmission oscillates around the value of the total transmission. The fifth $v_0 <\epsilon < v_1$ is a window where the transmission contains resonance peaks corresponding to the bound states associated with the double barrier tilting. The sixth zone $\epsilon>v_1$ contains oscillations, where the transmission converges to unity. Contrary to the case $v_0<v_1<v_2$, the results are modified, see Fig. \ref{fig1b}. Indeed, compared to Figs. \ref{fig1a}, the behaviors of some zones are completely reversed, like, for instance, the window zone. Fig. \ref{fig1c}  shows the behavior of the transmission for the case $v_0<v_2<v_1$, which is completely changed compared to Fig. \ref{fig1a} and \ref{fig1b}. However, if $L_1$ increases, the number of oscillations of transmission increases for $k_y<\epsilon<v_2-(k_y+\Delta_g/2)$ and $\epsilon> v_2-(k_y+\Delta_g/2)$ and decreases for$v_2-(k_y+\Delta_g/2)<\epsilon< v_2+(k_y+\Delta_g/2)$. Finally, we observe that the effect of the barrier width on transmission increases as long as $d_1$ increases, the minimal transmission of the intermediate zones $v_0<\epsilon<v_1$ decreases and the number of oscillations increases.

To illustrate the significance of our findings, we present two cases depicted in Fig. \ref{fig2}, which depend on the choice of barrier heights $(v_0, v_1, v_2)$. Specifically, Fig. \ref{fig2a} displays a double square barrier scenario with $v_2=20$ and $v_0\approx v_1=60$. The transmission in the Klein zone is not depicted, and the transmission oscillates around a minimum before behaving similarly to the results in Fig. \ref{fig1a}. However, the number of peaks decreases for double square barriers, while it increases for double barriers with tilting. 
Fig. \ref{fig2b} shows the case of a single square barrier with $v_0\approx v_1=v_2=40$. 
We note that the tilting and positioning of the scattering events within the barriers play a pivotal role in governing the peak of tunneling resistance. Specifically, the presence of a mid-barrier-embedded scatter causes a switch from a peak to a cusp when the incident energy reaches the Dirac point within a barrier. Moreover, the introduction of a continuously distributed scatter dampens constructive interference around the Dirac point, transforming a cusp into a peak in the tunneling resistance as the incident energy of electrons varies. In contrast to a single scatter, a continuous distribution within a barrier enhances unimpeded incoherent tunneling for head-on collisions and significantly suppresses skew collisions as the barrier tilting field increases \cite{Farhana}.


Let us explore the effects of introducing a time-varying potential with a sinusoidal oscillation of amplitude $U$ and frequency $\omega$ ($\alpha=\frac{U}{\omega}\neq 0$) in the intermediate region, where its height is oscillating around $v_2$. In Fig. \ref{fig3}, we show the transmission probabilities for the central band ($T_0$ in blue) and the first two sidebands ($T_{-1}$ in green and $T_1$ in red) versus the incident energy $\epsilon$, with the same as in Fig. \ref{fig1}. As anticipated, the transmissions are currently dispersed across both the central band and the sidebands. In addition, the utmost transmission via the oscillating barrier is dependent on the value of $\alpha$.


Fig. \ref{fig4} depicts the central transmission band $T_0$ (blue), and the first two sidebands $T_{-1}$ (red) and $T_{1}$ (green) as a function of the energy $\epsilon$. We consider a system with $L_{1}=1.5$, $L_{2}=2.5$, $\Delta_g=4$, $k_{y}=2$, $v_{0}=0$, $v_{1}=40$, and $v_{2}=60$. We investigate the effect of static and oscillating barriers with $\alpha=0$ (panel a), $\alpha=0.4$ (panel b), and $\alpha=0.98$ (panel c). Due to the sinusoidal and longitudinal vibrations of the time-oscillating barrier in the $x$-direction of Dirac fermions, with an amplitude of $u_1$ and frequency of $\omega$, the effective mass changes from $k_y$ to $k_y\pm\omega$. For small values of $\alpha$,  $T_0$ dominates, as shown in Fig. \ref{fig4a}, but as $\alpha$ increases,  $T_0$ decreases, and $T_{-1}$ as well as $T_{1}$ increase, as seen in Fig. \ref{fig4b}. In Fig. \ref{fig4c}, we see that transmissions of the two first sidebands ($l=\pm 1$) dominate. In conclusion, we notice that  increasing $\alpha$ leads to a decrease in $T_0$ and an increase in both $T_{\pm 1}$.


Fig. \ref{fig5} displays the transmission probabilities versus  potential $v_2$ with fixed parameters $L_1=0.4$, $L_2=0.8$, $\Delta_g=4$, $k_y=2$, $v_0=50$, $v_1=70$, and $\epsilon=60$. Here, the transmission probability $T_0$ (magenta) for the static barrier is shown in comparison to the transmission probabilities for the oscillating barrier with $\alpha\neq 0$. Specifically, the transmissions for the central band $T_0$ (blue) and the first two sidebands $T_{-1}$ (green) and $T_1$ (red) are shown for oscillating barriers with $\alpha=0.15$ (Fig. \ref{fig5a}), $\alpha=0.5$ (Fig. \ref{fig5b}), and $\alpha=0.98$ (Fig. \ref{fig5c}). We observe that the increase of $\alpha$ decreases  $T_0$ but increases  $T_{-1}$ and $T_{1}$.  Notably, the value of $\alpha$ plays a crucial role in determining the transmission probabilities for the sidebands. The two sidebands' transmissions are located near the axis of symmetry, where the central transmission strictly ranges between zero and one. This ensures that the sum of all transmissions does not exceed one, and both sidebands become symmetrical with respect to the opposite side of the axis of symmetry.

In Fig. \ref{fig6a}, we depict the transmission probabilities $T_{0,\pm 1}(\alpha=0.5)$ versus potential $v_1$ for  $L_{1}=1.5$, $L_{2}=2.5$, $\Delta_g=4$, $k_{y}=2$, $v_{0}=0$, $v_{2}=60$, and $\epsilon=30$. We see that  $T_0$  exhibits a sharp decline for $v_1>\epsilon+lw-(2k_y+\Delta_g)$,
reaching a relative minimum before oscillating and increasing again.
In Fig. \ref{fig6b}, we show the transmission probabilities versus energy gap $\Delta_g$ for $L_{1}=0.5$, $L_{2}=1.5$, $k_{y}=1$, $v_{0}=0$, $v_1=50$, $v_{2}=40$, and $\epsilon=15$. The maximum value of $T_0$ (magenta) is unity for $\alpha=0$. However, for $\alpha=0.5$, the maximum value of $T_0$ (blue) decreases, while the sideband transmissions $T_{-1}$ (green) and $T_{1}$ (red) increase. We observe that the sum of the three transmissions $T_0(\alpha=0.5)$, $T_{-1}(\alpha=0.5)$, and $T_{1}(\alpha=0.5)$ converges towards unity. Moreover, it should be noted that specific energy gaps completely block transmission. In fact, when the condition $\Delta_g> |\epsilon+lw- v_2|$ is satisfied, any incoming state is reflected.

\section{Conclusion}

We have studied the tunneling effect of electrons in gapped graphene passing through tilted double barriers and time-oscillating potential. More precisely, we have investigated how the transmission probabilities vary with respect to the incident energy, potential heights $v_2$ and $v_1$, energy gap, and incident angle. 
The results show that the tilting of the barriers and the position of the scatterers are crucial in tuning the peak of tunneling  when the incident energy approaches the Dirac point.
Moreover, the distributed scatterers effectively suppress the constructive interference around the Dirac point, which causes a change from a cusp to a peak in the tunneling resistance as a function of the incident  energy.


We have demonstrated that the oscillating barrier can play a crucial role in adjusting the transmission probabilities. Indeed, the time-varying potential induces additional sidebands in the transmission probability at energies of $\epsilon+l\hbar\omega$, arising from photon emission or absorption within the oscillating barrier. We have calculated the transmission probabilities for the central band and the first sidebands. Our findings indicate that  $T_{0}$ in the oscillating barrier decreases with increasing $\alpha$, whereas  $T_{-1}$ and $T_{1}$ increase. It is worth highlighting that the transmission probabilities for the sidebands through a graphene barrier tilted in a time-periodic potential are significantly influenced by the value of $\alpha$.

The exploration of effects in finite-width nanoribbons \cite{Roslyak2010, Oriekhov2018,Iurov2021} is highly pertinent to the present work. We believe that studying finite-width nanoribbons may provide valuable insights into how the bandgap properties of graphene structures can be tailored based on their geometric characteristics. Understanding how the width of nanoribbons affects electron transport properties is particularly relevant to our study. Exploring how geometric parameters, such as ribbon width or barrier spacing, influence electron transmission characteristics. These will provide a comprehensive understanding of the interplay between geometric features, external fields, and electron transmission properties. We look forward to delving into a more detailed discussion on these aspects in our future investigations. 

\section*{Data Availability Statement}
The data that support the findings of this study
are available on request from the corresponding author.

\appendix\label{TTTT}
\section{Transmission channels}\label{TTTT}

It is important to remember that as Dirac electrons travel through a region with a time-varying potential, they can transfer energy quanta to the oscillating field. This results in transitions from the central band to sidebands (channels) at energies $\epsilon\pm m\varpi$ $(m = 0, 1, \cdots)$
and therefore generates  transmission channels, which we have to determine.
Indeed, if we recognize that ${e^{imv_{F}\varpi t}}$ are mutually perpendicular, we can derive a set of equations based on the boundary conditions at $x=-L_2, -L_{1}, L_{1}, L_{2}$,  respectively, 
\begin{widetext}
	\begin{align}
		& \delta_{m,0} e^{-\textbf{\emph{i}}k_{m}
			L_{2}}+r_{m}e^{\textbf{\emph{i}}k_{m}
			L_{2}}=\sum^{l=\infty}_{l=-\infty}
		\left(a^{2}_{l}\eta_{1,l}^{+}(-L_2)
		+b^{2}_{l}\xi_{1,l}^{+}(-L_2)\right)
		\delta_{m,l}\lb{feq01}\\
		&  \delta_{m,0}z_{m}e^{-\textbf{\emph{i}}k_{m}
			L_{2}}-b_{m}^{1}\frac{1}{z_{m}}e^{\textbf{\emph{i}}k_{m} L_{2}}=
		\sum^{l=\infty}_{l=-\infty} \left(a^{2}_{l}
		\eta_{1,l}^{-}(-L_2)+b^{2}_{l}\xi_{1,l}^{-} (-L_2)\right)
		\delta_{m,l}\\
			&a^{2}_{l}\eta_{1,m}^{+}(-L_1)
			+b^{2}_{l}\xi_{1,m}^{+}(-L_1)=\sum^{l=\infty}_{l=-\infty}
			\left(a^{3}_{l}\lambda_{l}^{+}e^{-ik_{l}^{'}L_{1}}
			+b^{3}_{l}\lambda_{l}^{+}e^{ik_{l}^{'}L_{1}}\right)
			J_{m-l}\left(\alpha\right)\\
			& a^{2}_{l} \eta_{1,m}^{-}(-L_1)-b^{2}_{l}\xi_{1,m}^{-}
			(-L_1)=\sum^{l=\infty}_{l=-\infty}
			\left(a^{3}_{l}\lambda_{l}^{-}z_{l}^{'}e^{-ik_{l}^{'}L_{1}}-b^{3}_{l}\lambda_{l}^{-}\frac{1}{z_{l}^{'}}e^{ik_{l}^{'}L_{1}}\right)
			J_{m-l}\left(\alpha\right)\\
				&a^{4}_{l}\eta_{-1,m}^{+}(L_1)
				+b^{4}_{l}\xi_{-1,m}^{+}(L_1)=\sum^{l=\infty}_{l=-\infty}
				\left(a^{3}_{l}\lambda_{l}^{+}e^{ik_{l}^{'}L_{1}}
				+b^{3}_{l}\lambda_{l}^{+}e^{-ik_{l}^{'}L_{1}}\right)
				J_{m-l}\left(\alpha\right)\\
				& a^{4}_{l} \eta_{-1,m}^{-}(L_1)-b^{4}_{l}\xi_{-1,m}^{-}
				(L_1)=\sum^{l=\infty}_{l=-\infty}
				\left(a^{3}_{l}\lambda_{l}^{-}z_{l}^{'}e^{ik_{l}^{'}L_{1}}-b^{3}_{l}\lambda_{l}^{-}\frac{1}{z_{l}^{'}}e^{-ik_{l}^{'}L_{1}}\right)
				J_{m-l}\left(\alpha\right)\\
					& t_{m}e^{\textbf{\emph{i}}k_{m}
						L_{2}}+b_{m}^{5}e^{-\textbf{\emph{i}}k_{m}
						L_{2}}=\sum^{l=\infty}_{l=-\infty}
					\left(a^{4}_{l}\eta_{-1,l}^{+}(L_2)
					+b^{4}_{l}\xi_{-1,l}^{+}(L_2)\right)
					\delta_{m,l}\lb{feq01}\\
					&t_{m}z_{m}e^{\textbf{\emph{i}}k_{m}
						L_{2}}-b_{m}^{5}\frac{1}{z_{m}} e^{-\textbf{\emph{i}}k_{m} L_{2}}=
					\sum^{l=\infty}_{l=-\infty} \left(a^{4}_{l}
					\eta_{-1,l}^{-}(L_2)+b^{4}_{l}\xi_{-1,l}^{-} (L_2)\right)
					\delta_{m,l}. \lb{seq01}
				\end{align}
			\end{widetext}
				%
				We can write (\ref{feq01}-\ref{seq01})  in matrix form
				\begin{align}
					\begin{pmatrix}
						\Xi_{1} \\
						\Xi_{1}^{'} \\
					\end{pmatrix}%
					=
					\begin{pmatrix}
						{ \mathbb M_{11}} &{\mathbb M_{12}} \\
						{\mathbb M_{21}} &{ \mathbb M_{22}} \\
					\end{pmatrix}
					\begin{pmatrix}
						\Xi_{5} \\
						\Xi_{5}^{'}\\
					\end{pmatrix}={\mathbb M}
					\begin{pmatrix}
						\Xi_{5} \\
						\Xi_{5}^{'} \\
					\end{pmatrix}%
					 \lb{feqi}
				\end{align}
				where  ${\mathbb M}={\mathbb M(1,2)}
				\cdot {\mathbb M(2,3)} \cdot {\mathbb M(3,4)} \cdot {\mathbb
					M(4,5)}$ is the transfer matrix such that
				${\mathbb M(j,j+1)}$ connect the wave functions between the $j$-th and $(j+1)$-th regions
				\begin{align}
					&{\mathbb M(1,2)}=
					\begin{pmatrix}
						{\mathbb N_{1}^{+}}& {\mathbb N_{1}^{-}} \\
						{\mathbb L_{1}^{+}} &{\mathbb L_{1}^{-}} \\
					\end{pmatrix}^{-1}
					\begin{pmatrix}
						{\mathbb C_{1}^{+}} & {\mathbb G_{1}^{+}} \\
						{\mathbb C_{1}^{-}} & {\mathbb G_{1}^{-}} \\
					\end{pmatrix}%
					\\
					& {\mathbb M(2,3)}=
					\begin{pmatrix}
						{\mathbb Y_{1}^{+}}& {\mathbb Q_{1}^{+}} \\
						{\mathbb Y_{1}^{-}} &{\mathbb Q_{1}^{-}} \\
					\end{pmatrix}^{-1}
					\begin{pmatrix}
						{\mathbb D_{1}^{+}} & {\mathbb D_{1}^{-}} \\
						{\mathbb F_{1}^{+}} & {\mathbb F_{1}^{-}} \\
					\end{pmatrix}%
					\\
					&
					{\mathbb M(3,4)}=
					\begin{pmatrix}
						{\mathbb D_{-1}^{+}} & {\mathbb D_{-1}^{-}} \\
						{\mathbb F_{-1}^{+}} & {\mathbb F_{-1}^{-}} \\
					\end{pmatrix}^{-1}
					\begin{pmatrix}
						{\mathbb Y_{-1}^{+}}& {\mathbb Q_{-1}^{+}} \\
						{\mathbb Y_{-1}^{-}} &{\mathbb Q_{-1}^{-}} \\
					\end{pmatrix}%
					\\
					& {\mathbb M(4,5)}=
					\begin{pmatrix}
						{\mathbb C_{-1}^{+}} & {\mathbb G_{-1}^{+}} \\
						{\mathbb C_{-1}^{-}} & {\mathbb G_{-1}^{-}} \\
					\end{pmatrix}^{-1}
					\begin{pmatrix}
						{\mathbb N_{-1}^{+}}& {\mathbb N_{-1}^{-}} \\
						{\mathbb L_{-1}^{+}} &{\mathbb L_{-1}^{-}} \\
					\end{pmatrix}%
				\end{align}
				with the elements
				\begin{align}
					&\left({\mathbb N_{\gamma}^{\pm}}\right)_{m,l}=e^{\mp
						\gamma ik_{m}L_{2}}\delta_{m,l}\\
					&\left({\mathbb
						L_{\gamma}^{\pm}}\right)_{m,l}=\pm z_{m}^{\pm
						1}e^{\mp
						\gamma ik_{m}L_{2}}\delta_{m,l}\\
					& \left({\mathbb
						C_{\gamma}^{\pm}}\right)_{m,l}=\eta_{\gamma,m}^{\pm}(-\gamma L_2)\delta_{m,l}\\
					& \left({\mathbb G_{\gamma}^{\pm}}\right)_{m,l}=\xi_{\gamma,m}^{\pm}(-\gamma L_2)\delta_{m,l}\\
					& \left({\mathbb
						Y_{\gamma}^{\pm}}\right)_{m,l}=\eta_{\gamma,m}^{\pm}(-\gamma L_1)\delta_{m,l}\\
					& \left({\mathbb Q_{\gamma}^{\pm}}\right)_{m,l}=\xi_{\gamma,m}^{\pm}(-\gamma L_1)\delta_{m,l}\\
					& \left({\mathbb D_{\gamma}^{\pm}}\right)_{m,l}=\lambda^{+}e^{\mp
						\gamma ik_{l}^{'}L_{1}}J_{m-l}\left(\alpha\right)\\
					& \left({\mathbb
						F_{\gamma}^{\pm}}\right)_{m,l}=\pm\lambda^{-}(z_{l}^{'})^{\pm
						1}e^{\mp \gamma ik_{l}^{'}L_{1}}J_{m-l}\left(\alpha\right)
				\end{align}
				Here $\Xi_{1}=\{a_{m}^{1}\}=\{\delta_{m,0}\}$, $\tau \in\{1,2\}$, $\Xi_{1}^{'}=\{b_{m}^{1}\}
				=\{r_{m}\}$ represents the reflections, $\Xi_{5}^{'}=\{b_{m}^{5}\}$ denotes a null vector, 
				$\Xi_{5}=\{a_{m}^{5}\}=\{t_{m}\}$ represents the transmissions.
				%
				Consequently, we end up with 
				\begin{align}
					\Xi_{5}={ \mathbb M}_{11}^{-1} \cdot \Xi_{1}.
				\end{align}
				To find the minimum number of sidebands, we need to evaluate the strength of the oscillating potential, which can be expressed as $N>\alpha$ \cite{ahsan}. Consequently, we can simplify the infinite series for transmission by considering only a finite number of terms. Specifically, we can restrict the series to terms ranging from $-N$ to $N$ as reported in \cite{Mekkaoui1,Mekkaoui2,Mekkaoui3,Mekkaoui4,Mekkaoui5}. Therefore, we can express the series as follows:
				\begin{equation}\label{ttll}
					t_{-N+k}= { \mathbb M}_{11}^{-1}
					\left[k+1, N+1\right], \quad k=0, 1,  \cdots, 2N
				\end{equation}

				In getting the transmission  $T_{l}$ and reflection $R_{l}$ probabilities, we consider the current density $J$ associated with the present system that can be calculated as
				\begin{equation}
					J= v_{F}\psi^{\dagger}\sigma _{x}\psi
				\end{equation}
				giving rise to
				the incident $J_{\sf {inc,0}}$, transmitted $J_{{\sf {tra}},l}$, and reflected $ J_{{\sf {ref}},l}$, knowing that
				\begin{equation}
					T_{l}=\frac{ |J_{{\sf {tra}},l}|}{| J_{{\sf {inc}},0}|},\quad R_{l}=\frac{|J_{{\sf {ref}},l}|}{ |J_{\sf {inc,0}}|}.
				\end{equation}
				Recall that $T_{l}$ characterize the scattering of an electron with an incident energy  $\epsilon$ in region 1, into the sidebands with energies $\epsilon+l\varpi$ in region 5. 
				After some algebras, we obtain
				\begin{equation}\label{ttccs}
					T_{l}= \frac{k_{l}}{k_{0}}|t_{l}|^{2}, \quad R_{l}= \frac{k_{l}}{k_{0}}| r_{l}|^{2}.
				\end{equation}
				

\end{document}